\documentclass[prb,preprint,showpacs,preprintnumbers]{revtex4}
\usepackage{graphicx}
\usepackage{subfigure}
\usepackage{amssymb,amsmath}

\begin{document}

\bibliographystyle{apsrev}

\title{Interplay of chemical pressure and hydrogen insertion 
       effects in $ {\bf CeRhSn} $ from first principles} 

\author{A.\ F.\  Al Alam,$^a$
        S.\ F.\ Matar,$^a$\footnote{Corresponding author: matar$@$icmcb-bordeaux.cnrs.fr}
        N.\ Ouaini,$^b$ and  
        M.\ Nakhl$^b$}
\affiliation{$^a$ICMCB, CNRS, Universit\'e Bordeaux 1, 
             87 avenue du Docteur Albert Schweitzer, 
             33608 Pessac Cedex, France, \\
             $^b$Universit\'e Saint-Esprit de Kaslik, 
             Facult\'e des Sciences, 
             B.P. Jounieh, Lebanon.}

\date{\today}


\begin{abstract}
Investigations within the local spin density functional theory 
(LSDF) of the intermetallic hydride system $ {\rm CeRhSnH_x} $ 
were carried out for discrete model compositions in the range 
$ 0.33 \leq x_H \leq 1.33 $. The aim of this study is to assess  
the change of the cerium valence state in the neighborhood of 
the experimental hydride composition, $ {\rm CeRhSnH_{0.8}} $. 
In agreement with experiment, the analyses of the electronic and 
magnetic structures and of the chemical bonding properties point 
to trivalent cerium for $ 1 \leq x_H \leq 1.33 $. In contrast, 
for lower hydrogen amounts the hydride system stays in an 
intermediate-valent state for cerium, like in $ {\rm CeRhSn} $. 
The influence of the insertion of hydrogen is addressed from both 
the volume expansion and chemical bonding effects. The latter are 
found to have the main influence on the change of Ce valence 
character. Spin polarized calculations point to a finite magnetic 
moment carried by the Ce $ 4f $ states; its magnitude increases 
with $ x_H $ in the range $ 1 \leq x_H \leq 1.33 $.
\end{abstract}

\maketitle

\section{Introduction}
Equiatomic ternary alloys $ {\rm CeTX} $ (1:1:1), where T is a 
transition-metal of first, second and third period and X is a 
$ p $-element, form a rich family of intermetallic systems, due 
to the presence of X belonging to the $3A$, $4A$ and $5A$ main 
groups.\ \cite{adroja88,szytula94,riecken05,riecken07} These 
1:1:1 intermetallics have attracted considerable interest in the 
past two decades, due to a large variety of the magnetic and 
electrical properties. For instance, $ {\rm CeRhSn} $ is an 
intermediate-valent system,\ \cite{schmidt05} $ {\rm CePdSn} $ 
has a trivalent cerium,\ \cite{adroja88} while $ {\rm CeRuSn} $ 
comprises both trivalent and intermediate-valent cerium.\ \cite{riecken07} 
Furthermore, several 1:1:1 systems have the ability to absorb 
hydrogen.\ \cite{yartys02,chevalier06b,chevalier06} The hydrogenation 
of these compounds induces interesting physical properties relevant 
to the modification of the valence of cerium with respect to the 
initial 1:1:1 alloy system. Two effects can occur: (i) the expansion 
of the lattice by hydrogen intake, which should lead to an enhancement 
of the localization of the $4f$ states due to the reduced $ f $-$ f $ 
overlap; this effect occurs in the intermediate-valent $ {\rm CeNiIn} $, 
which on hydration transforms into a long-range magnetically ordered 
ferromagnet with trivalent Ce,\ \cite{chevalier02} (ii) the chemical 
interaction between the valence states of Ce, T, and X with H, which 
induces a decrease of the magnetic polarization and, eventually, 
a loss of magnetization such as in $ {\rm CeCoSiH} $\ \cite{chevmat} 
and $ {\rm CeCoGeH}$.\ \cite{chevmat1} Despite the numerous studies 
on 1:1:1 systems and corresponding hydrides, only little is known 
about the hydrogenation effects within the family of Sn based 
intermetallic systems. Recent experimental studies\ \cite{chevalier06} 
report the formation of a hydride $ {\rm CeRhSnH_{0.8}} $, within 
which a trivalent ground state of Ce was evidenced, but no detailed 
H positions were reported. Using the framework of the density 
functional theory (DFT),\ \cite{hohenberg64,kohn65,dftmethods} we 
examine in this work the electronic and magnetic properties of the 
experimental hydride composition within a range of discrete hydrogen 
amounts, for which full geometry optimizations were carried out. 
This allowed to assess the threshold for the change of Ce valence 
character upon hydrogenation found to be mainly induced by chemical 
interactions between the lattice constituents and H. 

\section{Crystal structures}
\begin{figure}[htbp]
\begin{center}
\includegraphics[width= 0.7\columnwidth ]{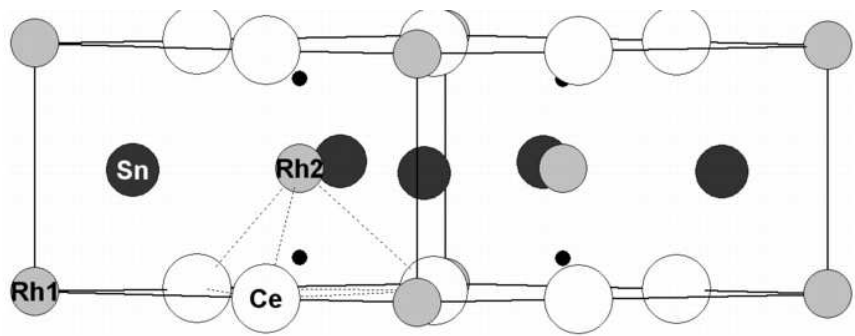}
\caption{The hexagonal crystal structure of $ {\rm CeRhSnH_{1.33}} $ 
         (space group $P\overline{6}2m$). Hydrogen atoms are 
          drawn as small black spheres. The tetrahedradral site 
          in which H are inserted is drawn in dotted lines.}	
\label{fig1}
\end{center}
\end{figure}

Like intermetallic  $ {\rm CeRhSn} $, the corresponding 
hydride systems $ {\rm CeRhSnH_x} $ crystallize in the 
$ {\rm ZrNiAl} $-type structure (space group $ P\bar{6}2m $).\ 
\cite{chevalier06} In this structure, Rh atoms are in {\it 1a} 
and {\it 2d} Wyckoff general positions, Rh1 (0, 0, 0) and Rh2 
($\frac {1}{3}$, $\frac {2}{3}$, $\frac {1}{2}$), while Ce and 
Sn are in 3-fold particular postions, {\it 3f} at (u$_{Ce}$, 0, 0) 
and {\it 3g} at ( u$_{Sn}$, 0, $\frac {1}{2}$). Since precise 
values for u$_{Ce}$ and u$_{Sn}$ are unknown for the hydride 
system, starting positions for the subsequent optimization were 
assumed as those of the 1:1:1 Sn based intermettallics, {\it i.e.}, 
u$_{Ce}$ = 0.414 and u$_{Sn}$ = 0.750.\ \cite{schmidt05} We note 
here that such positions are likely to change as a function of 
hydrogen contents as it will be detailed in the geometry optimization 
section. Within the hexagonal structure hydrogen is identified in 
a 4-fold {\it 4h} particular position at ( $\frac {1}{3}$, $\frac {2}{3}$, u$_H$) 
in a tetrahedral coordination with 3 Ce and one Rh2 atoms. From 
Yartys {\em et al.} works on $ {\rm CeNiIn} $ deuterides,\ \cite{yartys02} 
0.076$\leq$~u$_H$~$\leq$0.174 according to the amount of deuterium 
introduced, ranging from 0.48 up to 1.23. In the preliminary 
computations for $ {\rm CeRhSnH} $, u$_H$ was taken as 0.176. 
However, the complete filling of such arranged sites with hydrogen 
has been considered unlikely on the basis of the experimental 
results.\ \cite{chevalier06} In the present work, a discrete 
filling of these vacancies up to a full occupancy was performed. 
The fully saturated hydride with the stoichiometry $ {\rm Ce_3Rh_3Sn_3H_4} $ 
is sketched in Fig.\ \ref{fig1}. 

\section{Theoretical framework of computations}

\subsection{Computational methodology}
Two computational methods were used in the framework of density 
functional theory (DFT).\ \cite{hohenberg64,kohn65,dftmethods} 
A pseudo potential approach within the Vienna ab initio simulation 
package (VASP) code\ \cite{kresse96} was firstly used to optimize 
starting structures for different hydride compositions using 
projector augmented wave (PAW)\ \cite{blochl94,kresse99} potentials 
built within LDA\ \cite{perdew81} scheme. The calculations were 
converged at an energy cut-off of 301.01 \,eV for the plane-wave 
basis set with respect to the $ {\bf k} $-point integration with 
a starting mesh of 4*4*4 up to 8*8*8 for best convergence and 
relaxation to zero strains. The Brillouin-zone integrals were 
approximated using a special $ {\bf k} $-point sampling. Further,  
this method allows for a first insight into the charge density of 
the hydride system. 

The all-electron calculations are based on density-functional theory 
and the local-density approximation (LDA) as parametrized according 
to Vosko, Wilk, and Nusair.\ \cite{vosko} They were performed using 
the scalar-relativistic implementation of the augmented spherical 
wave (ASW) method (see Refs.\ \onlinecite{wkg,aswbook} and references 
therein). In the ASW method, the wave function is expanded in atom-centered
augmented spherical waves, which are Hankel functions and numerical
solutions of Schr\"odinger's equation, respectively, outside and 
insidethe so-called augmentation spheres. In order to optimize the 
basis set, additional augmented spherical waves were placed at carefully 
selectedinterstitial sites (IS). The choice of these sites as well 
as the augmentation radii were automatically determined using the 
sphere-geometry optimization algorithm.\ \cite{sgo} Self-consistency 
was achieved by a highly efficient algorithm for convergence acceleration.\ 
\cite{mixpap} The Brillouin zone integrations were performed using 
the linear tetrahedron method with up to 1088 {\bf k}-points within 
the irreducible wedge.\ \cite{blochl94,aswbook} The efficiency of 
this method in treating magnetism and chemical bonding properties in 
transition-metal, lanthanide and actinide compounds has been well 
demonstrated in recent years.\ \cite{chev,eyert04,mat3} 

\subsection{Assessment of chemical bonding properties}
To extract more information about the nature of the interactions 
between the atomic constituents from electronic  structure calculations, 
the crystal orbital overlap population (COOP) \cite{hoffmann87} 
or the crystal orbital Hamiltonian population (COHP) \cite{dronskowski93} 
may be employed. While both the COOP and COHP approaches provide 
a qualitative description of the bonding, nonbonding, and antibonding 
interactions between two atoms, the COOP description in some cases 
exaggerates the magnitude of antibonding states. A slight refinement 
was recently proposed in form of the so-called covalent bond energy 
ECOV criterion, which combines the COHP and COOP to calculate 
quantities independent of the particular choice of the potential 
zero.\ \cite{bester01} In the present work, the ECOV criterion was 
used for the chemical bonding analysis. In the plots, negative, 
positive and zero magnitudes of the unitless ECOV are indicative 
of bonding, antibonding, and nonbonding interactions respectively. 

\section{Geometry optimized discrete compositions of hydrides}
In as far as no structural determination for the particular Ce, 
Sn and H postions were available, trends in cell volumes and ground 
state crystal structures for the different amounts of H were 
necessary in the first place. The equilibrium structures were 
obtained starting from  $ {\rm CeRhSn} $\ \cite{schmidt05} structural 
setup for u$_{Ce}$ and u$_{Sn}$ internal parameters for Ce and Sn 
positions. As for H, our starting guesses were based on the experimental 
study performed by Yartys {\em et al.}\ \cite{yartys02} on $ {\rm CeNiInD_x} $ 
structures. An initial input for the lattice $a$ constant for the 
different compositions was done on the basis of the experimental 
increase of $a$ between the 1:1:1 alloy and the experimental 
hydride composition, {\it i.e.}, $\frac {\Delta a}{a} = 1.5 \%$. 
\cite{chevalier06} The hexagonal $\frac{c}{a}$ ratio of 0.547 of 
$ {\rm CeRhSnH_{0.8}}$ taken from the experimental data\ \cite{chevalier06} 
was preserved throughout the calculations, thus allowing for an 
isotropic volume evolution of the different structures upon the 
intake of discrete amounts of hydrogen. After that, computations were 
carried out for the hydride systems with different amounts of H. 
>From the results, the hexagonal symmetry was preserved for the 
optimized geometries of all systems. The new values for u$_{Ce}$, 
u$_{Sn}$ and u$_H$ internal parameters are given in table\ \ref{tab1}. 
As it can be expected, u$_{Ce}$, u$_{Sn}$ are close to starting 
values. As for u$_H$ all results are in agreement with starting 
ones except for $ {\rm CeRhSnH_{0.33}} $ where the small value of 0.028 
is obtained. This value is close to zero, which allows assuming 
that H within $ {\rm CeRhSnH_{0.33}} $ is relaxed into the general 2-fold 
position {\it 2c} at ($\frac {1}{3}$, $\frac {2}{3}$, 0). According
to the total energies values extracted from the VASP calculations, 
which are shown in table\ \ref{tab1}, stabilization is proportional 
to the amount of the absorbed hydrogen. Further, the values of $\Delta{E}$, 
which is the relative energy of the hydride system with respect 
to  $ {\rm CeRhSn} $, confirm this tendency. However, to investigate 
the origin of this stability, the energetic amount $\frac {n}{2} E_{H_2}$ 
was substracted from $E$. Here $n$ is an interger ranging from 0 to 
4, and $E_{H_2}$ is the energy of H$_2$ (-0.489 \,$Ryd$) considered 
as the sum of the energy of two widely separate hydrogen atoms and 
the dissociation energy of H$_2$ (0.329 \,$Ryd$/mol H$_2$) as given 
by experiment.\ \cite{atkins83} As presentend in table\ \ref{tab1} 
these values point to the chemical bonding of H with other atomic 
constituents as responsible of this stability. Furthermore, we 
notice that ($E-\frac {n}{2}E_{H_2}$) value for $ {\rm CeRhSnH_{0.66}} $ 
is smaller than that of  $ {\rm CeRhSn} $, pointing to a difficulty 
to obtain such a hydride experimentally. 

\begin{figure}[htbp]
\begin{center}
\includegraphics[width=\columnwidth ]{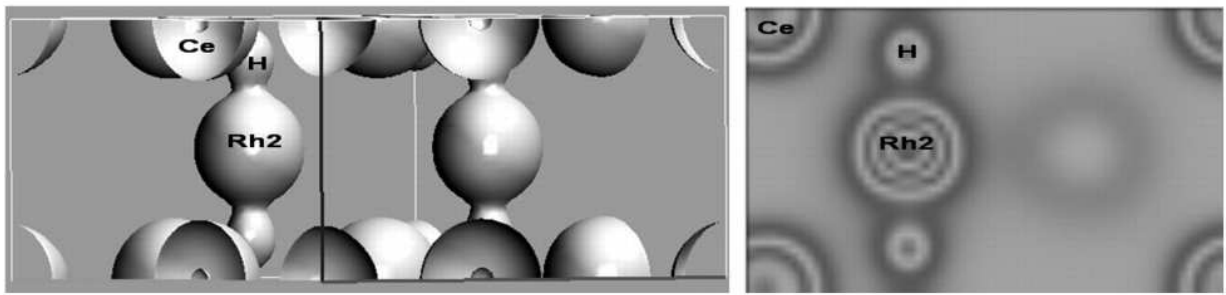}
\caption{Charge density plot for $ {\rm CeRhSnH_{1.33}} $: the 
         isosurface and volume slice are sketched on the left 
         and right hand sides respectively, both are drawn with c hexagonal 
		 axis along the paper sheet. (Plots with VMD software.\ \cite{vmd}).}	
\label{fig2}
\end{center}
\end{figure}

One can also analyze the electron charge density around the chemical 
species which allows determining the amount of localization of 
electrons. Fig.\ \ref{fig2} shows the isosurface as well as the 
volume slice of the charge density both divided by the unit cell 
volume within the saturated hydride system. This illustration 
points to the strong bonding between Rh2 and H along the $c$-axis. 

\section{All-electron ASW computations}
The crystal parameters provided by the VASP geometry optimizations 
(table \ref{tab1}) were used for the input of all-electron calculations. 
For $ {\rm CeRhSnH_x} $ compositions lower than that of the 
saturated hydride system $ {\rm CeRhSnH_{1.33}} $, the positions of the 
lacking hydrogen atoms within the {\it 4h} interstices were considered as 
interstitial sites (IS) where augmented spherical waves were placed. 
The resulting breaking of initial crystal symmetry was accounted for 
from the differentiation of the crystal constituents in the calculations. 
In a first step the calculations were carried out assuming non-magnetic 
configurations (non spin polarized NSP), meaning that spin degeneracy 
was enforced for all species. However, note that such a configuration 
is different from that of a paramagnet, which could be simulated for 
instance by a supercell entering random spin orientations over the different 
magnetic sites. Subsequent  spin polarized calculations (spin-only) 
lead to an implicit long-range ferromagnetic ordering. In order to 
provide a model for a antiferromagnetic (AF) ground state of the 
hydride system and since neutron diffraction data were not available, 
we have constructed a double unit cell along the $c$-axis. This 
provides one possible model of a long-range AF spin structure which 
should be validated as a ground state configuration from the relative 
energies of the band theoretical calculations. On the other hand, 
other sets of computations were performed for hydrogen-free models 
within ASW, starting from the previous four hydride systems with an 
additional model at the same volume of the experimental hydride.\ 
\cite{chevalier06} This procedure evaluates the manner in which the 
volume expansion affects the magnetic behavior of cerium. 

\subsection{Spin degenerate calculations}
\subsubsection{Density of states}
The characteristic features of the site-projected DOS (PDOS) for 
the saturated hydride system $ {\rm CeRhSnH_{1.33}} $ plotted in 
Fig.\ \ref{fig3b} enable the discussion of the bonding in the 
$ {\rm CeRhSnH_x} $ models. Here and in all following figures, 
energies are referred to the Fermi level $ {\rm E_F} $. In the 
PDOS three energy regions can be distinguished. The first one, 
from -12 to -5 \,eV, comprises $5s$ (Sn) and $1s$ (H) states. 
Then from -6 \,eV up to $ {\rm E_F} $ shows $5p$ (Sn) states which 
hybridize with $4d$ (Rh) states and itinerant Ce states. Finally, 
rather localized Ce $ 4f $ states are found at and above $ {\rm E_F} $. 
The resulting high density of states at $ {\rm E_F} $ is indicative 
for an instability of the system in a degenerate spin configuration 
as is discussed next. In comparison with previous results obtained 
for  $ {\rm CeRhSn} $ (Fig.\ \ref{fig3a}),\ \cite{Matar07} 
Fig.\ \ref{fig3b} shows that on hydrogenation  new states are formed 
around -12 to -8 \,eV. Hence the valence band (VB) is shifted to 
lower energies and $ {\rm E_F} $ is pushed slightly upwards due to 
the additional electrons, resulting in a larger width of the VB 
with respect to $ {\rm CeRhSn} $. 

\subsubsection{Analysis of the DOS within Stoner theory}
In as far as $4f$ (Ce) states were treated as band states by our 
calculations, the Stoner theory of band ferromagnetism\ \cite{dftmethods} 
can be applied to address the spin polarization. The total energy 
of the spin system results from the exchange and kinetic energies 
counted from a non-magnetic state. Formulating the problem at zero 
temperature, one can express the total energy as 
$ E= \frac{1}{2} [\frac{m^2}{n({\rm E_F})}][1- {\rm I} n({\rm E_F})] $. 
Here I is the Stoner exchange integral, which is an atomic 
quantity that can be derived from spin polarized calculations.\ 
\cite{janak} $n({\rm E_F})$ is the PDOS value for a given species 
at the Fermi level in the non-magnetic state. The product I$n({\rm E_F})$ 
from the expression above provides a criterion for the stability 
of the spin system. The change from a non-magnetic configuration 
towards spin polarization is favorable when I$n({\rm E_F})\geq$1. 
The system then stabilizes through a gain of energy due to exchange. 
From Ref.\ \onlinecite{matar2000}, I(Ce-$4f$)$\sim 0.02\,Ryd $ and the 
computed $n({\rm E_F})$ values of Ce ($4f$) for all the model systems 
are given in table\ \ref{tab2}. The calculated values for I$ n({\rm E_F}) $ 
of 0.53, 0.41, 1.05, and 2.30 for $ {\rm CeRhSnH_{0.33}}$, $ {\rm CeRhSnH_{0.66}} $, 
$ {\rm CeRhSnH} $, and $ {\rm CeRhSnH_{1.33}} $, respectively, point 
to a magnetic instability of for the latter two hydride systems. 
This prediction for the behavior of the valence of Ce will be checked 
within the spin-polarized calculations for further confirmation, wherby 
finite magnetic moment are expected to be carried by $4f$ (Ce) states 
of $ {\rm CeRhSnH} $ and $ {\rm CeRhSnH_{1.33}} $. Also, table\ \ref{tab2} 
gives the values of Stoner product for hydrogen-free models. Among the 
computed hydrogen-free models, $ {\rm CeRhSnH} $ and $ {\rm CeRhSnH_{1.33}} $ 
are the only ones, which are susceptible to a magnetic disorder, but their 
clearly weaker I$n({\rm E_F})$ values of 0.93 and 0.96, with respect to 
their pure hydride models analogues, emphasizes the contribution 
of hydrogen chemical bonding to the arising of the trivalent 
character of cerium over the volume expansion. Lastly, an additional 
hydrogen-free model at the same volume of the experimental hydride\ 
\cite{chevalier06} was computed. The calculated value of its Stoner 
product is equal to 0.76, which points to a non favorable trend 
for magnetic instability. This is a further confirmation of the 
dominant role of hydrogen chemical bonding.   

\subsubsection{Chemical bonding}
Chemical bonding properties can be readily addressed on the basis 
of the spin-degenerate calculations. This is due to the fact that 
the spin-polarized bands, to a large degree, result from the 
spin-degenerate bands by a rigid spin splitting. The hydride models 
$ {\rm CeRhSnH_x} $ have the same bonding trends of CeRhSn obtained from 
the calculations of Matar {\em et al.} (Fig.\ \ref{fig4a}) using 
the ECOV criterion.\ \cite{Matar07} A visual inspection of Fig.\ 
\ref{fig4b} shows that the dominant interactions within the VB 
result from the Ce-Rh2 and Rh1-Sn bonds. This is concomitant with 
the interatomic distances given in table\ \ref{tab2}. Whereas the 
smaller the distance is the stronger is the interaction, Ce-Rh2 
as well as Rh1-Sn have the shortest separations with respect to 
other Ce-T and T-Sn distances. However, compared to each other, 
the separation of Ce-Rh2 has the larger value with respect to that 
of Rh1-Sn suggesting a stronger Rh1-Sn bond. Nevertheless, Ce-Rh2 
remains the most stabilizing contribution, which is due to the Rh2 
interaction with hydrogen. In fact, Fig.\ \ref{fig2} 
shows the continuous isosurface shape of charge density for 
the two hydrogens surrounding Rh2. This 
is confirmed by the interatomic distances given in table\ \ref{tab2} 
for all $ {\rm CeRhSnH_x} $ models, where the Rh2-H separation is clearly 
the shortest.   

\subsection{Spin polarized configurations}
>From the NSP calculations and their analysis within the Stoner 
mean field theory of band ferromagnetism, it has been established 
that the hydride system is unstable in such a configuration for 
H content close to the experiment, {\it i.e.} with $x_H$ =1 up to 
1.333. Consequently, spin polarized calculations were carried out, 
assuming implictly a hypothetic ferromagnetic order. This is 
done by initially allowing for two different spin occupations, 
then the charges and the magnetic moments are self-consistently 
converged. The relative energies differences ($\Delta{E}$), with 
respect to the spin-degenerate energy of  $ {\rm CeRhSn} $, 
for spin-degenerate and spin-polarized calculations given in table\ \ref{tab2}, 
slightly favor the ferromagnetic state. The experimental finding 
suggested a trivalent character of Ce within $ {\rm CeRhSnH_{0.8}} $. 
Theoretical non-magnetic computations of the discrete intake of 
hydrogen within the hydride systems $ {\rm CeRhSnH_x} $ also reported 
a magnetic instability threshold at the hydride composition of three 
hydrogen. Spin-degenerate calculations confirmed these tendencies by 
identifying finite spinonly magnetic moments of 0.173 and 0.377 $\mu_{B}$ 
carried by $4f$ (Ce) states for both $ {\rm CeRhSnH} $ and $ {\rm CeRhSnH_{1.33}} $ 
respectively. These results are illustrated at Fig.\ \ref{fig5} showing 
the site and spin projected density of states of $ {\rm CeRhSnH_{1.33}} $. 
The exchange splitting is observed for cerium. The main bonding 
characteristics follow the discussion above of the non-magnetic PDOS. 

Lastly, in order to check for the nature of the magnetic ground 
state, AF calculations were carried out using a supercell built 
from two simple cells along the $c$-axis. These two structures 
were used to distinguish between the up- and down-spin atoms. At 
self-consistency, the energy difference ($\Delta{E}=E_{Ferro}-E_{AF}=-10^{-2}\,Ryd$) 
extracted from the spin-polarized calculation for both ferro- 
and antiferromagnetic state favors the ferromagnetic ordering, 
thus pointing to a ferromagnetic ground state. Further experimental
investigations are underway. 

\section{Conclusion}
In this work we have undertaken a theoretical investigation of 
the effects of the insertion of hydrogen into  $ {\rm CeRhSn} $ on 
the valence state character of cerium. Hydride models $ {\rm CeRhSnH_x} $, 
are computed herein to estimate both volume and hydrogen chemical 
bonding interplay within the change of the valence character for 
Ce. The properties of these models were addressed both by spin-degenerate 
and spin-polarized LSDF based calculations. Analyses of the electronic 
structures and of the chemical bonding reveal different types of 
chemical bonds due to the nature of the hydrogen tetrahedral 
interstices and their environment with mainly Rh2 atoms. Furthermore, 
a local magnetic moment is expected only for the cerium site with 
trivalent character whose arising threshold is for 1 H per formula 
unit. Compared to the experimental finding of 2.4 hydrogen atoms, 
this result is rather satisfying. Spin-polarized calculations lead 
to a finite moment showing up only at the trivalent Ce site. 

\section{Acknowledgments}
We acknowledge discussions with Priv. Doz. Dr Volker Eyert 
(University of Augsburg, Germany). 

Computational facilities 
were provided by the M3PEC-M\'esocentre of the University Bordeaux 1, financed by the 
``Conseil R\'egional d'Aquitaine'' and the French Ministry of 
Research and Technology. 

\newpage
\begin{table}[htbp]
\begin{tabular}{lccccc}
\hline\hline
                                          &$ {\rm CeRhSn} $ &$ {\rm CeRhSnH_{0.33}} $ &$ {\rm CeRhSnH_{0.66}} $ &$ {\rm CeRhSnH} $ &$ {\rm CeRhSnH_{1.33}} $ \\ 
\hline  
a ($ \AA $)                               &7.448            &7.465                    &7.532                    &7.598             &7.664 \\
c ($ \AA $)                               &4.080            &4.136                    &4.136                    &4.136             &4.136 \\
Volume ($ \AA^3 $)                        &196              &197                      &202                      &208               &213 \\
E ($Ryd$)                                 &-4.35            &-4.62                    &-4.83                    &-5.12             &-5.36 \\
$E-\frac {n}{2}E_{H_2}$ ($Ryd$)           &-4.35            &-4.37                    &-4.34                    &-4.39             &-4.38 \\
$\Delta{E}$ ($Ryd$)                       &0                &-0.489                   &-0.489                   &-0.771            &-1.016 \\
u$_{Ce}$                                  &-                &0.412                    &0.412                    &0.404             &0.395 \\
u$_{Sn}$                                  &-                &0.743                    &0.744                    &0.744             &0.743 \\
u$_H$                                     &-                &0.028                    &0.130                    &0.131             &0.139 \\
\hline\hline
\end{tabular}
\caption{$ {\rm CeRhSn} $ and $ {\rm CeRhSnH_x} $ model systems: 
         u$_{Ce}$, u$_{Sn}$ and u$_H$ are particular positions for 
         Ce, Sn and H refined from geometry optimization calculations 
         using VASP code.}
\label{tab1}
\end{table}

\begin{table}[htbp]
\begin{tabular}{lccccc}
\hline\hline
                   &$ {\rm CeRhSnH_x} $ &$ {\rm CeRhSnH_{0.33}} $ &$ {\rm CeRhSnH_{0.66}} $ &$ {\rm CeRhSnH} $ &$ {\rm CeRhSnH_{1.33}} $ \\ 
\hline  
$ \Delta{E} $      &0                   &-0.607                   &-1.736                   &-2.684            &-3.794 \\
                   &{\it 0}             &{\it -0.608}             &{\it -1.736}             &{\it -2.685}      &{\it -3.797} \\
I$n({\rm E_F})$    &0.69                &0.53                     &0.41                     &1.05              &2.30 \\
                   &{\it 0.69}          &{\it 0.81}               &{\it 0.69}               &{\it 0.93}        &{\it 0.96} \\
$ d_{Ce-Rh1} $     &3.084               &3.095                    &3.121                    &3.148             &3.174 \\
$ d_{Ce-Rh2} $     &3.031               &3.036                    &3.063                    &3.089             &3.116 \\
$ d_{Ce-Sn} $      &3.227               &3.232                    &3.264                    &3.290             &3.317 \\
                   &3.375               &3.386                    &3.412                    &3.444             &3.475 \\
$ d_{Rh1-Rh2} $    &4.756               &4.772                    &4.814                    &4.856             &4.898 \\
$ d_{Rh1-Sn} $     &2.761               &2.767                    &2.793                    &2.814             &2.841 \\
$ d_{Rh2-Sn} $     &2.846               &2.851                    &2.878                    &2.904             &2.925 \\
$ d_{Ce-H} $       &-                   &2.306                    &2.328                    &2.343             &2.365 \\
$ d_{Rh1-H} $      &-                   &4.338                    &4.380                    &4.417             &4.454 \\
$ d_{Rh2-H} $      &-                   &1.534                    &1.545                    &1.561             &1.571 \\
                   &-                   &4.576                    &4.613                    &4.655             &4.697 \\
$ d_{Sn-H} $       &-                   &3.237                    &3.264                    &3.296             &3.322 \\
\hline\hline
\end{tabular}
\caption{$ {\rm CeRhSn} $ and $ {\rm CeRhSnH_x} $ model systems: 
         interatomic distances are given in $ \AA $. Given in units 
         of $ Ryd $, $\Delta{E}$ represents the energy of the hydride 
         with respect to the scalar-relativistic energy of  
         $ {\rm CeRhSn} $ ( $E_0$=-118836.5904\,$Ryd$) both resulting 
         from spin-degenerate calculations. The values in italic font 
         for $\Delta{E}$ correspond to the energy of the hydride for 
         the spin-polarized computations relative to $E_0$. As for 
         I$n({\rm E_F})$, the italic font results are related to 
         hydrogen-free hydride models.}
\label{tab2}
\end{table}

\newpage

\begin{figure}[htbp]
\begin{center}
\subfigure[~]{\includegraphics[width=0.49\columnwidth ]{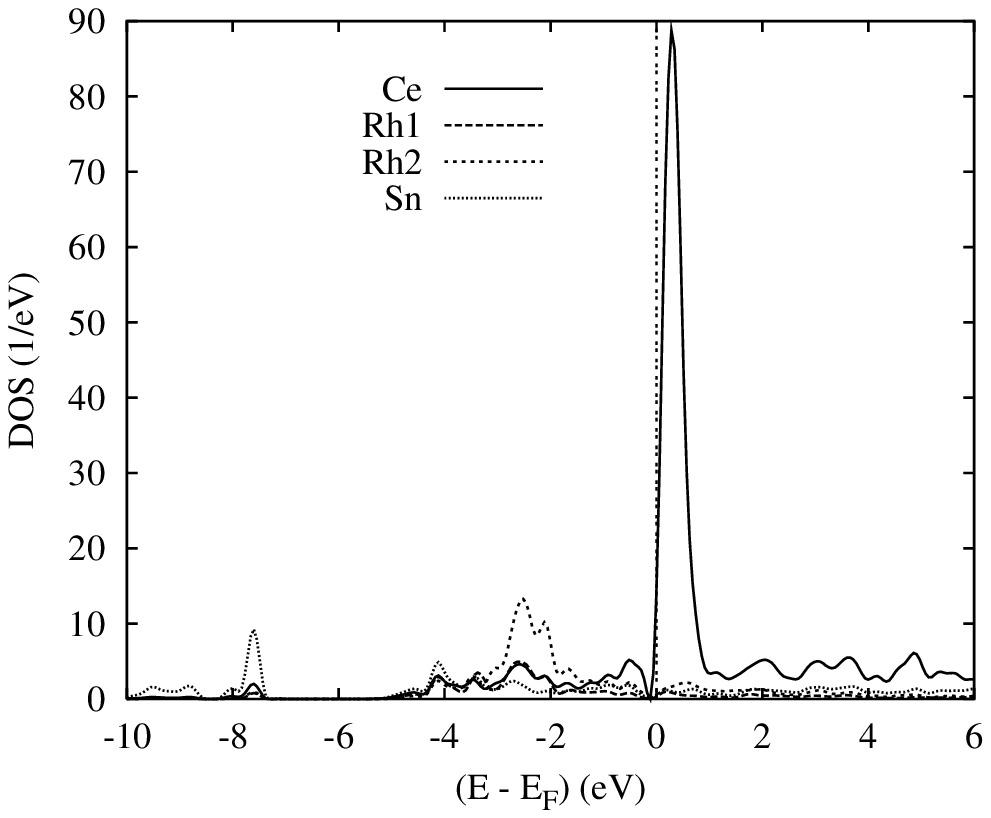}\label{fig3a}}
\subfigure[~]{\includegraphics[width=0.49\columnwidth ]{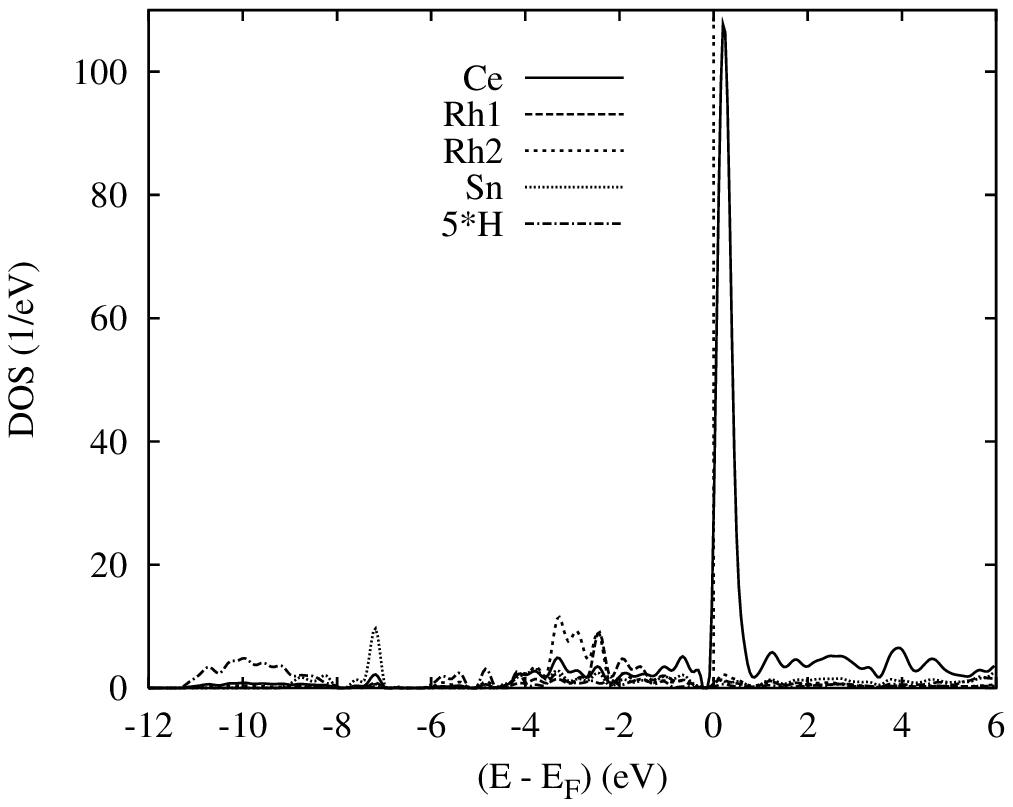}\label{fig3b}}
\caption{Non-magnetic site projected DOS for: $ {\rm CeRhSn} $ (a) and 
         $ {\rm CeRhSnH_{1.33}} $, for the sake of clear presentation 
         hydrogen PDOS are multiplied by 10 (b).}
\end{center}
\end{figure}

\begin{figure}[htbp]
\begin{center}
\subfigure[~]{\includegraphics[width=0.49\columnwidth]{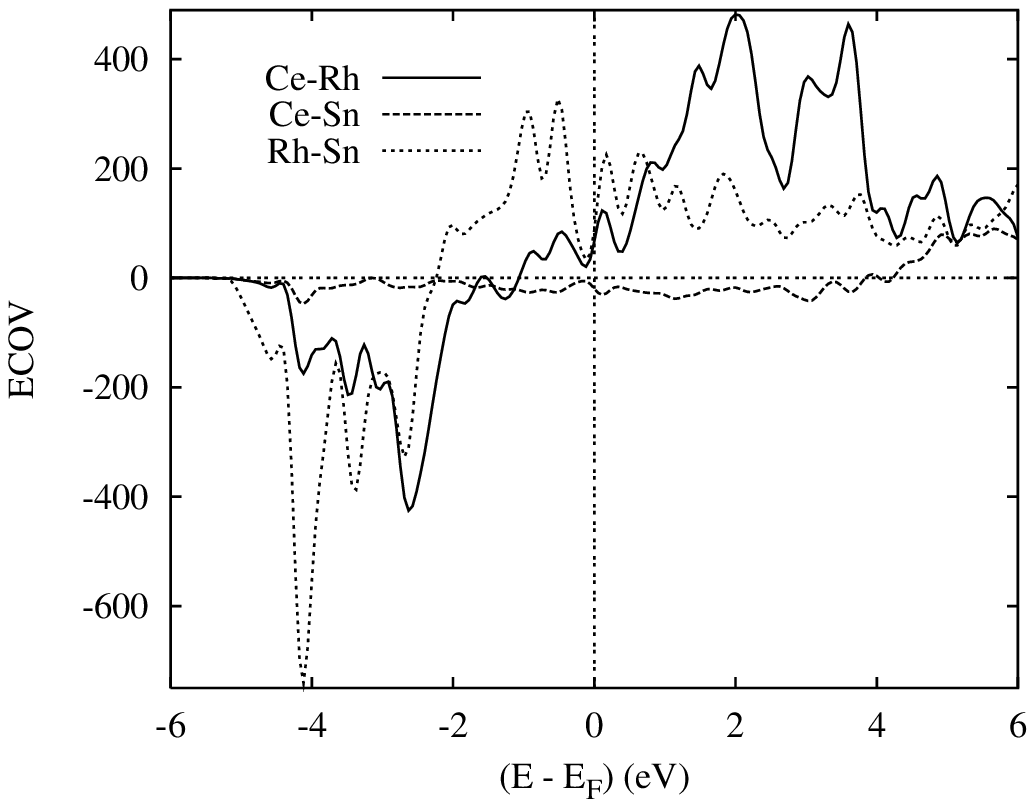}\label{fig4a}}
\subfigure[~]{\includegraphics[width=0.49\columnwidth]{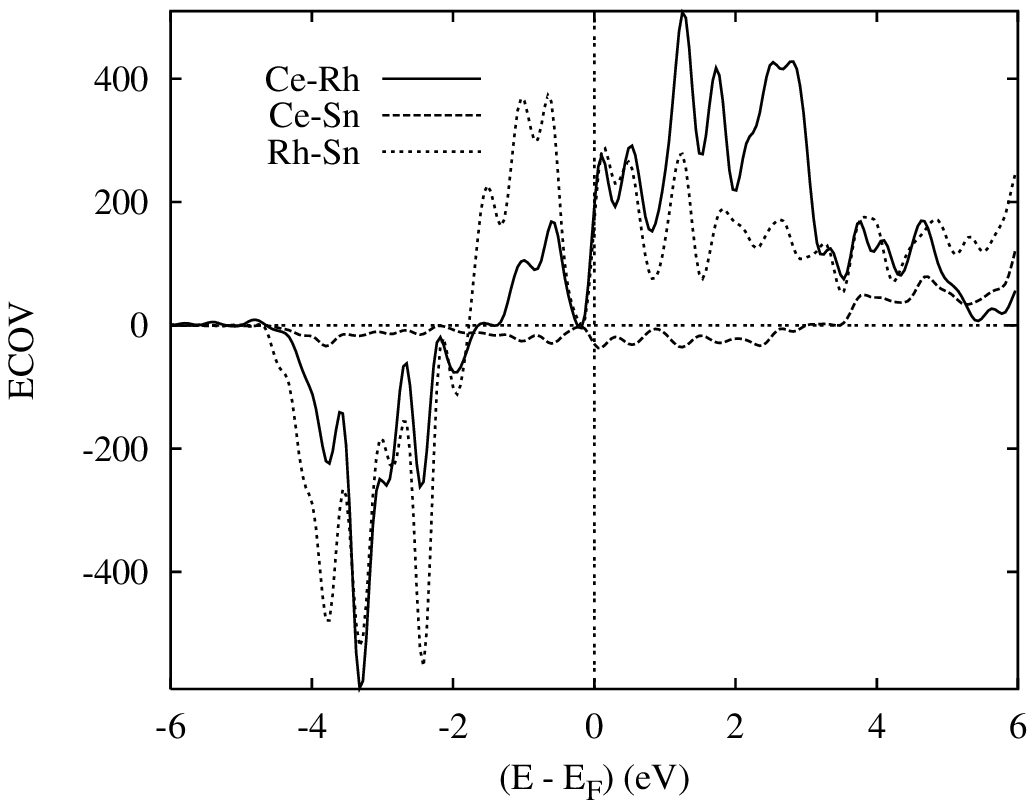}\label{fig4b}}
\caption{Chemical bonding with ECOV criterion for: 
         NM-$ {\rm CeRhSn} $ (a), and $ NM-{\rm CeRhSnH_{1.33}} $ (b).}
\end{center}
\end{figure} 

\begin{figure}[htbp]
\begin{center}
\includegraphics[width=0.8\columnwidth]{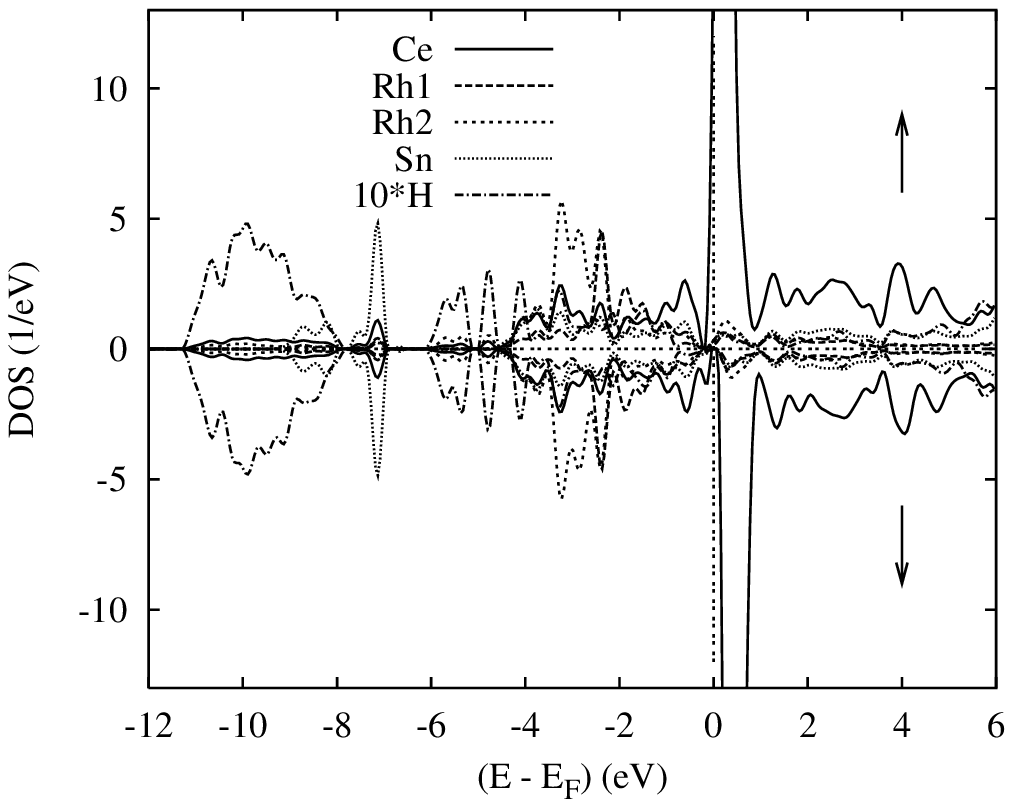}
\caption{Site and spin projected DOS of $ {\rm CeRhSnH_{1.33}} $ 
         in the ferromagnetic state.}
\label{fig5}
\end{center}
\end{figure}

 

\begin{thebibliography}{999}
\bibitem{adroja88} 
D.\ T.\ Adroja, S.\ K.\ Malik, B.\ D.\ Padalia, and R.\ Vijayaraghavan, 
Solid State Commun.\ {\bf 66}, 1201 (1988).

\bibitem{szytula94} 
A.\ Szytu\l a, J.\ Leciejewicz, 
{\em Handbook of Crystal Structures and Magnetic Properties of Rare 
Earth Intermetallics}, CRC Press, Boca Raton, Florida (1994).

\bibitem{riecken05} 
J.\ F.\ Riecken, G.\ Heymann, T.\ Soltner, R.\ -D.\ Hoffmann, 
H.\ Huppertz, D.\ Johrendt, and R.\ P\"ottgen, 
Z.\ Naturforsch.\ {\bf 60b}, 821 (2005).

\bibitem{riecken07}
J.\ F.\ Riecken, W.\ Hermes, B.\ Chevalier, R.\ -D.\ Hoffmann, 
F.\ M.\ Schappacher, and R.\ P\"ottgen, 
Z.\ Anorg.\ Allg.\ Chem.\ {\bf 633}, 1094 (2007).

\bibitem{schmidt05} 
T.\ Schmidt, D.\ Johrendt, C.\ P.\ Sebastian, R.\ P\"ottgen, 
K.\ \L \c{a}tka, and R.\ Kmie\'c, 
Z.\ Naturforsch.\ {\bf 60b}, 1036 (2005).

\bibitem{yartys02}
V.\ A.\ Yartys, R.\ V.\ Denys, B.\ C.\ Hauback, H.\ Fjellv\aa g, 
I.\ I.\ Bulyk, A.\ B.\ Riabov, and Ya.\ M. Kalychak,   
J.\ Alloys Comp.\ {\bf 330-332}, 132 (2002).

\bibitem{chevalier06b}
B.\ Chevalier, A.\ Wattiaux, and J.\ -L.\ Bobet, 
J.\ Phys.\ : Condens.\ Matter {\bf 18}, 1743 (2006).

\bibitem{chevalier06}
B.\ Chevalier, C.\ P.\ Sebastian, and R.\ P\"ottgen, 
Solid State Sci.\ {\bf 8}, 1000 (2006).

\bibitem{chevalier02}
B.\ Chevalier, M.\ L.\ Kahn, J.\ -L.\ Bobet, M.\ Pasturel, 
and J.\ Etourneau,
J.\ Phys.\ : Condens.\ Matter {\bf 14}, L365 (2002).

\bibitem{chevmat} 
B.\ Chevalier, and S.\ F.\ Matar, 
Phys.\ Rev.\ B {\bf 70}, 174408 (2004).
 
\bibitem{chevmat1} 
B.\ Chevalier, S.\ F.\ Matar, J.\ Sanchez Marcos, and J.\ Rodriguez Fernandez, 
Physica B: Condensed Matter {\bf 378-380}, 795 (2006).

\bibitem{hohenberg64}
P.\ Hohenberg, and W.\ Kohn, 
Phys.\ Rev.\ B {\bf 136}, 864 (1964). 

\bibitem{kohn65} 
W.\ Kohn, and L.\ J.\ Sham, 
Phys.\ Rev.\ A {\bf 140}, 1133 (1965).

\bibitem{dftmethods} 
J.\ K\"ubler, V.\ Eyert, 
{\em Electronic structure calculations} in 
{\em Materials Science and Technology. Vol. 3A: Electronic and Magnetic 
Properties of Metals and Ceramics}, Part I.\ Volume Editor K.\ H.\ J.\ 
Buschow (VCH, Verlag, Weinheim), 1-145 (1992).

\bibitem{kresse96}
G.\ Kresse, and J.\ Furthm\"uller, 
Phys.\ Rev.\ B {\bf 54}, 11169 (1996).

\bibitem{blochl94} 
P.\ E.\ Bl\"ochl, 
Phys.\ Rev.\ B {\bf 50}, 17953 (1994).

\bibitem{kresse99} 
G.\ Kresse, and J.\ Joubert, 
Phys.\ Rev.\ B {\bf 59}, 1758 (1999).

\bibitem{perdew81} 
J.\ P.\ Perdew, and A.\ Zunger, 
Phys.\ Rev.\ B {\bf 23} 5048 (1981).

\bibitem{vosko} 
S.\ H.\ Vosko, L.\ Wilk, and M.\ Nusair. 
Can.\ J.\ Phys.\ {\bf 58}, 1200 (1980).

\bibitem{wkg} A.\ R.\ Williams, J.\ K\"ubler and C.\ D.\ Gelatt,
Phys.\ Rev.\ B {\bf 19}, 6094 (1979). 

\bibitem{aswbook}
V.\ Eyert, 
{\em The Augmented Spherical Wave Method -- A Comprehensive Treatment}, 
Lect.\ Notes Phys.\ {\bf 719} (Springer, Berlin Heidelberg 2007).

\bibitem{sgo} 
V.\ Eyert and K.\ -H.\ H\"ock, 
Phys.\ Rev.\ B {\bf 57}, 12727 (1998).

\bibitem{mixpap}
V.\ Eyert, 
J.\ Comput.\ Phys.\ {\bf 124}, 271 (1996).
\bibitem{chev} 
B.\ Chevalier, and S.\ F.\ Matar, 
Phys.\ Rev.\ B {\bf 70}, 174408 (2004).

\bibitem{eyert04}
V.\ Eyert, C.\ Laschinger, T.\ Kopp, and R.\ Fr\'esard, 
Chem.\ Phys.\ Lett.\ {\bf 385}, 249 (2004).

\bibitem{mat3} 
S.\ F.\ Matar, E.\ Gaudin, B.\ Chevalier, and R.\ P\"ottgen, 
Solid State Sci.\ {\bf 9}, 274 (2007).

\bibitem{hoffmann87} 
R.\ Hoffmann, 
Angew.\ Chem.\ Int.\ Ed.\ Engl.\ {\bf 26}, 846 (1987). 

\bibitem{dronskowski93} 
R.\ Dronskowski and P.\ E.\ Bl\"{o}chl, 
J.\ Phys.\ Chem.\ {\bf 97}, 8617 (1993). 

\bibitem{bester01} 
G.\ Bester and M.\ F\"ahnle, 
J.\ Phys: Condens.\ Matter {\bf 13}, 11541 (2001).

\bibitem{atkins83} 
P.\ W.\ Atkins, 
{\em Molecular Quantum Mechanics}, 2$^{nd}$ ed.\, 
(Oxford University Press, Oxford,1983), p.\ 257.

\bibitem{Matar07}
S.\ F.\ Matar, J.\ F.\ Riecken, B.\ Chevalier, R.\ P\"ottgen, 
A.\ F.\ Al Alam, and V. Eyert, 
Phys.\ Rev.\ B {\bf 76}, 174434 (2007).

\bibitem{janak} 
J.\ F.\ Janak, 
Phys.\ Rev.\ B {\bf 16}, 255 (1977).

\bibitem{matar2000} 
S.\ F.\ Matar and A.\ Mavromaras,
J.\ Solid State Chem.\ {\bf 149}, 449 (2000). 

\bibitem{vmd}
W.\ Humphrey, A.\ Dalke, and K.\ Schulten, 
J.\ Molec.\ Graphics {\bf 14}, 33 (1996).
\end{thebibliography}
\end{document}